ISSN : 2229-4333(Print) | ISSN : 0976-8491(Online)  IJCST Vol. 2, Issue 3, September 2011# Secure Transmission of Password Using Speech Watermarking

[1]Rupa Patel, [2]Urmila Shrawankar, [3]Dr. V.M Thakare
[1,2]G.H.Raisoni College of Engineering, Nagpur, India
[3]S.G.B. Amravati University, Amravati, Maharashtra, India.## Abstract
Internet is one of the most valuable resources for information communication and retrievals. Most multimedia signals today are in digital formats. The digital data can be duplicated and edited with great ease which has led to a need for data integrity and protection of digital data. The security requirements such as integrity or data authentication can be met by implementing security measures using digital watermarking techniques. In this paper a blind speech watermarking algorithm that embeds the watermark signal data in the musical (sequence) host signal by using frequency masking is used. A different logarithmic approach is proposed. In this regard a logarithmic function is first applied to watermark data. Then the transformed signal is embedded to the converted version of host signal which is obtained by applying Fast Fourier transform method. Finally using inverse Fast Fourier Transform and antilogarithmic function watermark signal is retrieved.

## Keywords
Digital Watermarking, logarithmic, speech signals, data hiding technique, frequency masking## I. Introduction
Digital watermarking is a technique to embed information into the underlying data. A digital watermark can be created from user or transaction specific information, which can be embedded in the speech. The embedded information can then be detected and verified at the receiver side. Most of the multimedia digital signals are easy to manipulate that led to a need for security of these signals. Using digital watermarking techniques the security requirements such as data integrity, data authentication can be met.

### A. Watermark model
An embedder has two inputs, one is the watermark message that can contain information such as password, copyright, authorship etc. and the other input is the cover signal (host signals or carrier) that can be speech signals, video, image etc.

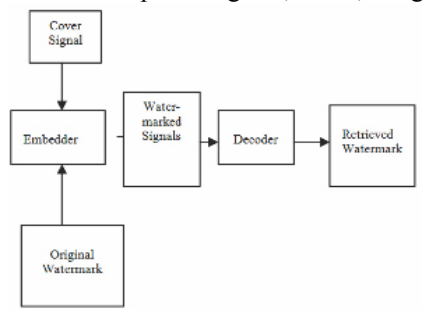

Fig.1: General watermark model

The output of the watermark embedder is the watermarked signal, which cannot be perceptually discriminated from the host signal. The watermarked signal is then presented to the watermark detector. The detector determines whether the watermark is present in the tested multimedia signal, and if so, what message is encoded in it.

### B. Requirement for digital watermarking
In order to be an effective watermarking system it should possess the following characteristics

1. The watermark should not affect the quality of the original signal, thus it should be inaudible to human ears.

2. It should be robust to resist common signal processing manipulations.

3. The cover signal should be larger than the watermark signal.

4. The watermark should only be detected by authorized person.

5. The speed of embedding of watermark is important in real time applications where the embedding is done on continuous signals.

6. It is recommended to have unique watermarks to different files to help make the technique more useful.

7. One requirement is that the invisibility of the watermark should not be compromised by multiple applications.

8. Watermark detection should be done without referencing the original signals

## II. Related work
In recent years, digital watermarking techniques achieved significant progress [11]. Early methods considered watermarking as a purely additive process with potential spectral shaping of the watermark signal or an additive embedding of the watermark in a transform domain of the host signal. This type of system was used in many practical implementations and proved to be highly robust, but suffered from the inherent interference between the watermark and the host signal. Most watermarking methods use a perceptual model to determine the permissible amount of embedding-induced distortion. Many audio watermarking algorithms use auditory masking, and most often frequency masking, as the perceptual model for watermark embedding [7]. Several algorithms for the embedding and extraction of watermarks in audio sequences have been presented. All of the developed algorithms take advantage of the perceptual properties of the human auditory system (HAS) in order to add a watermark into a host signal in a perceptually transparent manner. [16]. In general, speech may be considered as a class of one dimensional signal—and of audio signals, in particular. Thus, watermarking techniques that have been previously proposed for the audio domain may be applied to the speech watermarking problem. Nevertheless, this approach fails to recognize and exploit the domain specific characteristics for improved performance. Similarly, it fails to recognize and counter malicious attacks that

www.ijcst.com   INTERNATIONAL JOURNAL OF COMPUTER SCIENCE AND TECHNOLOGY   315

are not applicable to general class of audio signals, yet encountered frequently in the speech domain [12].In speech watermarking, spread spectrum type system is used to be method of choice for robust embedding [11].Only a few watermarking methods have been proposed specifically for the speech domain. In [13], Wu et al. propose a QIM (quantization index modulation) technique that operates on the DFT (discrete Fourier transform) coefficients. The method is tuned for the speech domain by its exponential scaling property, which targets the psychoacoustic masking functions and band-pass characteristics. QIM methods embed the information by requantizing the signals, in generalization some methods modulate the speech signal or one of its parameter according to the watermark data [3] Auditory masking describes the psycho-acoustical the principle of auditory masking is exploited either by varying the quantization step size or embedding strength in one way or the other, or by removing masked components and inserting a watermark signal in replacement principle that some sounds are not perceived in the temporal or spectral vicinity of other sounds  In particular, frequency masking (or simultaneous masking) describes the effect that a signal is not in the presence of a simultaneous louder masker signal at nearby frequencies[14]. There is no perceptual difference between different realizations of the Gaussian excitation signal for non-voiced speech .It is possible to exchange the white Gaussian excitation signal by a white Gaussian data signal that carries the watermark information. The signal thus forms a hidden data channel within the speech signal [15].

## III. Data hiding algorithm

In this section we propose the watermarking algorithm for data hiding. In the watermarking-communications mapping, the process of watermarking is seen as a transmission channel through which the watermark message is being sent, with the non voiced host signal being a part of that channel. Frequency masking approach has been used to embed the watermark signal components into high frequency subband of the host signal. We are using the long-known fact that, in particular for non-voiced speech and blocks of short duration, the ear is insensitive to the signal's phase.

The steps of watermarking embedding algorithm are expressed as follows
- Watermark embedding
- Watermark extraction

### A. Watermark embedding

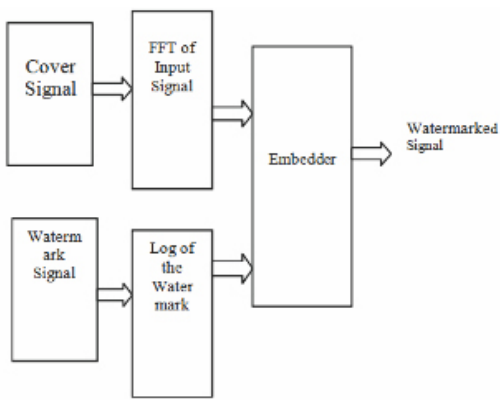

Fig. 2: Watermark embedding process

Watermark embedding can be summarized as

1. The important step in the processing of the signal is to obtain a frequency spectrum of the input signal. The information in the frequency spectrum is used for extracting features such as high frequency components. One method to obtain a frequency spectrum is to apply a Fast Fourier Transform (FFT). The digital input signal undergoes a transformation that outputs a collection of FFT coefficients termed "host vectors" or "host signals" or "cover signal".
2. The noise has been removed using wiener filter [5] and then watermark signal is transformed using logarithmic function.
3. Determine the center of the density of high frequency input signal. Then, the watermark embedding is performed on high frequency components of the host signal using frequency masking method to form watermarked signal.
4. After the added pattern is embedded, the watermarked work is usually distorted during watermark attacks. We model the distortions of the watermarked signal as added noise.

### B. Watermark extraction

Once a signal has been watermarked, next step is to deal with the extraction of the watermark sequence. However, because a digitally watermarked signal is obtained by invisibly hiding information into the host signal. The password/secret message is recovered using an appropriate decoding process. The challenge is to ensure that the watermarked signal is perceptually indistinguishable from the original and that the message be recoverable.

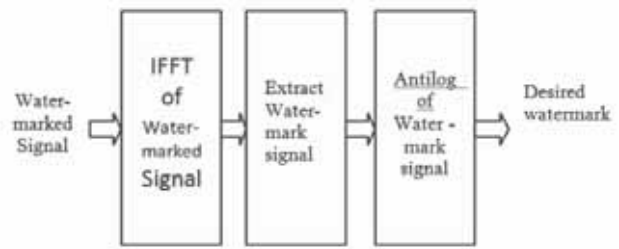

Fig.3 : Watermark extraction process

Watermark Extraction can be summarized as
1. The digital watermarked signal undergoes a transformation that outputs a collection of coefficients (Inverse Fast Fourier Transform i.e. IFFT).
2. The high frequency components of the watermarked signal are extracted.
3. Then, antilog of the extracted watermark is performed to form recover watermarked signal.

## IV. Experimental setup and results

We have considered non-voiced signal i.e. instrumental sound file as a host or cover signal. The sample rate of the host signal was found to be 22050Hz.
%read the input wave sound
[input_sound fs] = wavread ('8.wav');
The original host sound signal is shown in Fig. 4.

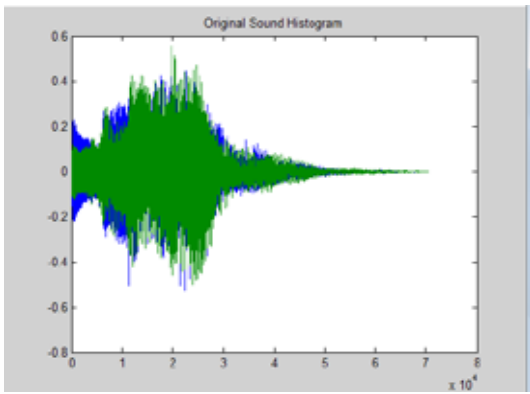

Fig. 4: Original cover signal

The password 'FIVE' is the secret message that we are embedding in the host signal. The sampling frequency of the watermark signal was recorded to be 8000 Hz.
% Watermark file
[watermark_sound fs1]= wavread ('five.wav');
The histogram of the watermark signal is shown by the fig.5.

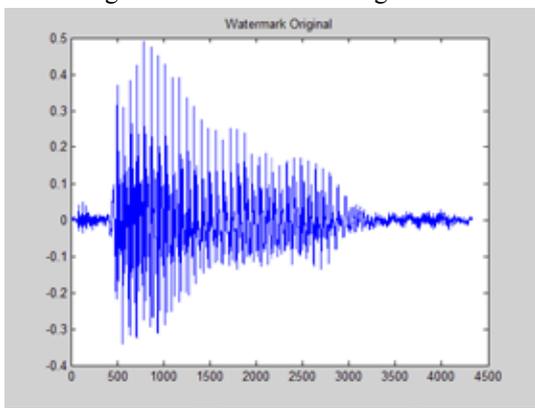

Fig.5: Original watermark signal

The Fast Fourier transformation is applied on the original sound signal. When FFT function is performed, a separate vector is created which becomes the frequency value. The logarithmic/exponential function is applied on watermark signal before embedding it with host signal.
%find the fft of the input
fft_sound = fft2 (sound);
log_watermark = exp (watermark);
index = 1; %apply watermarking for count=length (fft_sound)-WatermarkLength+1:length(fft_sound)
   fft_sound (count) = log_watermark (index);
   index = index + 1; end

The watermarked signal thus obtained is depicted in Fig. as a reconstructed sound signal.

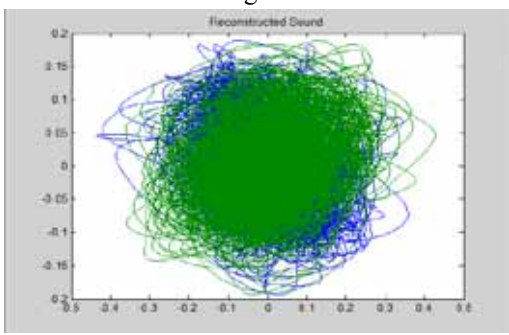

Fig. 6: Reconstructed sound signals

To retrieve original watermark signal from the watermarked signal, inverse Fourier transform of reconstructed sound is done and then antilog of extracted watermark signal is performed.
% Watermark Extraction
%Get the fft of reconstructed sound
output_fft = real(fft2(ifft2(reconstructed_sound)));

%Fetch the watermark from this sound
index = 1;
watermark = zeros ([1 WatermarkLength]);
for count=length(received_sound)-WatermarkLength+1:length(received_sound)
watermark (index) = output_fft (count);
index = index + 1;
end

%Find the log value of this watermark
watermark = log (watermark);

The Fig. 7 shows the histogram of the retrieved watermark signal.

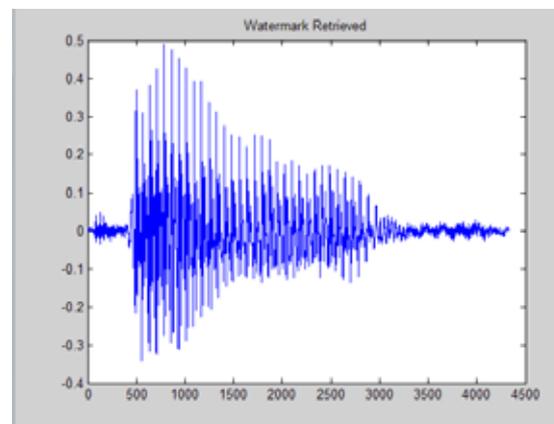

Fig. 7: Retrieved watermark signal

We found that embedding strategy is satisfactory since we were not able to guess whether the signal was watermarked or not during the subjective testing. Next step consists of the extraction of the watermark sequence. Result shows that the watermark is perfectly extracted. From the observations, it is to be noted that the embedded audio signal is imperceptible.

### V. Conclusion
In this paper we have implemented a watermarking technique for secure transmission of password within a cover signal based on logarithmic approach, taking features of Human Auditory System and the signal processing theories. Although experimental results shows that the watermark is likely not hearable and information carried by the host signal i.e. watermark are retrievable still there remains a scope of improvement. Further work can be carried by considering issues close to the robustness of the methods against various attacks.

### References

[1] Dr. Vipula Singh."Digital Watermarking - A Tutorial", Cyber Journals: Multidisciplinary Journals in Science and Technology, Journal of Selected Areas in Telecommunications (JSAT),January edition 2011, pp 10-21.
[2] Roland Kwitt, et.al. "Lightweight Detection of Additive waterminking in the DWT–Domain ", IEEE Transactions on Image Processing, Vol 20, issue 2, 2010, pp 474 – 484,

[3] Nima K Kalantari , S. Ahadi."Logarithmic Quantization Index Modulation : A Perceptually Better Way to Embed Data within a cover Signal", IEEE International Conference on Acoustics, Speech and Signal Processing, ICASSSP 2009,19-24 April, 2009, pp 1433-1436.

[4] Mohamed Waleed Fakhr."A Novel Data Hiding Technique for Speech Signals with High Robustness", IEEE International Symposium on Signal Processing and Information Technology, 2007, pp 379-384

[5] Urmila Shrawankar, Dr. V M Thakare. "Noise Estimation and Noise Removal Techniques for speech Recognition in Adverse Environment", (Springer-IIP2010), Manchester, UK, October 13-16, 2010

[6] S.Saraswathi." Speech Authentication based on Audio Watermarking", International Journal of Information Technology, Vol. 16 No. 1, 2010

[7] Konard Hofbauer, et.al. " Speech Watermarking for Analog Flat Fading Bandpass Channels" , IEEE Transactions on Audio, Speech, and Language Processing, Vol 17,Issue 8,2009, pp 1624-1637.

[8] Dr. Hana'a M. Salman."A Content-Based Authentication Using Digital Speech Data", Eng. & Tech., Vol.25, No.10, 2007

[9] Duhamel, P., M. Vetterli."Fast Fourier Transforms: A Tutorial Review and a State of the Art," Signal Processing, Vol. 19, 1990, pp. 259-299.

[10] C. Wu and C. Jay Kuo. "Comparison of two speech content authentication approaches," in Proc. SPIE: Security and Watermarking of Multimedia Contents IV, E. J. Delp and P. W. Wong, Eds., vol. 4675, Jan. 2002, pp. 158–169.

[11] I. Cox, M. Miller, J. Bloom, J. Fridrich, . Kalker." Digital Watermarking and Steganography" , 2nd ed. San Mateo, CA:Morgan Kaufmann, 2007

[12] Mehmet Celik, et.al. "Pitch and Duration Modification for Speech Watermarking", IEEE International Conference on Acoustics, Speech and Signal Processings,Vol 2, 2005, pp 17-20.

[13] C. Wu and C. Jay Kuo. "Fragile speech watermarking based on exponential scale quantization for tamper detection," in Proc. IEEE Intl. Conf. Acoustics Speech and Sig. Proc., May 2002, pp 3305–3308.

[14] S. Chen and H. Leung. "Concurrent data transmission through PSTN by CDMA," in Proceedings of the IEEE International Symposium on Circuits and Systems (ISCAS), Island of Kos, Greece, May 2006, pp. 3001–3004.

[15] G. Kubin, B. S. Atal, W. B. Kleijn."Performance of noise excitation for unvoiced speech," in Proceedings of the IEEE Workshop on Speech Coding for Telecommunications, Saint-Adele, Canada, Oct. 1993, pp. 35–36.

[16] C. Nedeljko, "Algorithms For Audio Watermarking And Steganography",Academic Dissertation, University of Oulu, public discussion in Kuusamonsali , Linnanmaa, Finland, 2004.

[17] X. Wang, H. Zhao. "A Blind Audio Watermarking Robust Against Synchronization Attacks," CIS 2005, Part II, LNAI 3802, 2005, pp. 617-622.

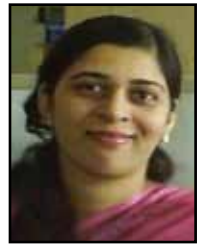

Rupa Patel did her MCA from Nagpur University and is pursuing M.Tech in Computer Science and Engineering.Her area of interest are Speech recognition, Green Computing and Data Mining.

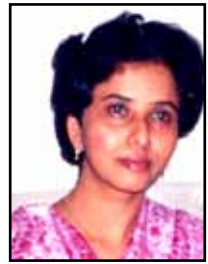

Urmila Shrawankar did her M.Tech (Computer Science) and is a Research Scholar in Computer Science and Engineering at Amravati University. Her areas of interest are Operating System, Human Computer Interaction, Speech Recognition System. She has 44 publications in International Journals and conferences. Received funding support from Dept. of Science and Technology (DST) for this research work.